# Title: Model-Free Idealization: Adaptive Integrated Approach for Idealization of Ion Channel Currents (AI2)

Madoka Sato[1], Masanori Hariyama[2*], Komiya Maki[3], Kae Suzuki[4,5], Yuzuru Tozawa[4], Hideaki Yamamoto[3], Ayumi Hirano-Iwata[1,3,6*]

[1]Graduate School of Biomedical Engineering, 2-1-1 Katahira, Aoba-ku, Sendai-shi, Miyagi 980-8577, Japan

[2]Graduate School of Information Sciences, Tohoku University, 6-3-09, Aoba, Aramaki, Aoba, Sendai, 980-8579, Japan

[3]Laboratory for Nanoelectronics and Spintronics, Research Institute of Electrical Communication, Tohoku University, 2-1-1 Katahira, Aoba-ku, Sendai-shi, Miyagi 980-8577, Japan

[4]Graduate School of Science and Engineering, Saitama University, 255 Shimo-Okubo, Sakura-ku, Saitama-shi, Saitama 338-8570, Japan

[5]Epsilon Molecular Engineering, Inc., Rm208, Research bldg. of Open Innovation Center in Saitama university, 255 Shimo-okubo, Sakura-ku, Saitama city, Saitama 338-8570 Japan

[6]Advanced Institute for Materials Research (WPI-AIMR), Tohoku University, 2-1-1 Katahira, Aoba-ku, Sendai-shi, Miyagi 980-8577, Japan

*Correspondence: masanori.hariyama.c3@tohoku.ac.jp, ayumi.hirano.a5@tohoku.ac.jp

**ABSTRACT (no more than 300 words)**

Single-channel electrophysiological recordings provide insights into transmembrane ion permeation and channel gating mechanisms. The first step in the analysis of the recorded currents involves an "idealization" process, in which noisy raw data are classified into two discrete levels corresponding to the open and closed states of channels. This provides valuable information on the gating kinetics of ion channels. However, the idealization step is often challenging in cases of currents with poor signal to-noise ratios (SNR) and baseline drifts, especially when the gating model of the target channel is not identified. We report herein on a highly robust model-free idealization method for achieving this goal. The algorithm, called AI2 (Adaptive Integrated Approach for the Idealization of Ion Channel Currents), is composed of Kalman filter and Gaussian Mixture Model (GMM) clustering and functions without user input. AI2 automatically determines the noise reduction setting based on the degree of separation between the open and closed levels. We validated the method on pseudo channel current datasets which contain either computed or experimentally recorded noise. The AI2 algorithm was then tested on actual experimental data for biological channels including gramicidin A, a voltage-gated sodium channel, and other unidentified channels. We compared the idealization results with those obtained by the conventional methods, including the 50%threshold-crossing method.

## Introduction

Ion channels are transmembrane proteins that allow specific ions to pass through the highly insulative cell membrane (1, 2). Ion flow through such ion channels is essential for the proper functioning of neural, muscle, and cardiac systems through the generation of an action potential and synaptic transmission (3). Upon opening the channel pore, ion channels allow ions to flow passively along their electrochemical potential gradient. The resulting ionic currents, ion channel currents, contain valuable information regarding channel gating kinetics and inherent ion conductance (4). However, most single-channel currents are typically very small, in the pA range, and contain background noise (5). Therefore, the first step in analyzing ion channel currents is a so-called ”idealization” process in which noisy raw data are classified into two states, yielding time-series data composed of open and closed states.

There are two main types of idealization: model-based and model-free methods. Single-channel analysis based on Hidden Markov Models (HMMs) has been extensively studied and several applications, such as QuB (6, 7), HJCFIT (8), SPARTAN (9), and Deep-Channel (10) have been developed. Although these model-based idealization methods are strong tools, it is necessary to specify gating kinetics and rate constants of the target molecules in advance (11-15). This represents a serious limitation to the analysis, especially if the ion channel kinetics are complex and the gating model is not easily estimated.

Model-free idealization is a method that does not require *a priori* knowledge of likely gating models, such as the 50%-threshold-crossing method (16-18), J-SMURF (19), and Minimum Description Length (MDL) (20). These types of model-free methods can be used to study various channels without adopting a gating model and setting rate constants for the target channels. Instead, these methods require user dependent input that can significantly affect the idealization results. J-SMURF and MDL require user inputs that involve adjusting the level of false positive detections (20, 21). The most widely used 50%-threshold-crossing approach also requires a cutoff frequency to reduce high-frequency noise before binarization (17). However, the wide range of channel events, with open times ranging from milliseconds to seconds, hampers appropriate settings for the low-pass filter. Researchers have been forced to determine the cutoff frequency empirically, which can cause important signals to be eliminated and/or insufficient noise reduction.

This report describes an automatic idealization method called Adaptive Integrated Approach for Idealization of ion channel currents: AI2. AI2 iteratively combines Kalman-filter-based noise reduction and Gaussian-Mixture-Model (GMM)based clustering and then determines whether or not the level of noise reduction is appropriate. A Kalman filter can estimate the unmeasurable system state from the noisy data by minimizing the estimated error covariance (22), and GMM clustering can stochastically classify data points into Gaussian distribution components (23). The level of noise reduction is automatically determined by the results of GMM fitting. AI2 includes a function that allows baseline drifts to be removed based on the RANSAC approach (24) which robustly fits a linear baseline model to faulty data. We assessed the performance of AI2 using datasets that simulate ion channel currents with different signal-to-noise ratios (SNRs) and noise sources. The datasets contained either electrophysiological recording noise or white Gaussian noise. We also applied AI2 to the idealization of ion channel currents for three different types of ion channels, including unidentified ones. These idealization results were compared with those obtained by conventional methods, including 50%-threshold-crossing.

## Materials and Methods

### The structure of AI2

Users first input time-series current data into AI2. AI2 then removes baseline drift (detrend) and performs denoising and clustering to yield output binary signals. The process of detrend is skipped for the analysis of pseudo-channel-currents datasets which contained no baseline drifts.

### Denoise-clustering algorithm of AI2

A Kalman filter provides the optimal estimates of an unknown time series signal when the observed measurements are given over time. In order to handle the uncertainty of the system efficiently, the Kalman filter takes the variance of system noise into account. The level of noise reduction is determined by a parameter κ which represents the initial value of error covariance. When κ becomes larger, the level of noise reduction becomes lower, which could result in noisy signals. Conversely, when κ becomes smaller, the level of noise reduction becomes higher, which may result in too much filtering. Therefore, it is important to find the appropriate κ value for the Kalman filter.

AI2 allows an automatic search for the appropriate κ from $1.0\times10^{-1}$ to $1.0\times10^{-4}$ based on the distribution of the signal. For each κ, two parameters $S_1$ and $S_2$ (see below) are calculated until the κ that minimizes $S_1$ or $S_2$ is obtained. After noise reduction using the determined κ, the distribution of the currents would be bimodal with two peaks that are distant from each other and/or are distinguished, leading to successful automatic clustering (Fig. 1). The parameters, $S_1$ and $S_2$, are defined as follows:

$$S_1 = \frac{\sigma_1+\sigma_2}{|\mu_1-\mu_2|} \quad (Eq.1)$$

$$S_2 = \frac{PD_{th}}{PD_{small\ peak}} \quad (Eq.2)$$

where $\sigma_1$ and $\sigma_2$ are the standard deviations of $peak_1$ and $peak_2$, respectively, of the twocomponent gaussian mixture model (GMM); $\mu_1$ and $\mu_2$ are the mean values for $peak_1$ and $peak_2$, respectively; $PD_{small\ peak}$ is the probability density value of the fitted GMM at the smaller peak, and $PD_{th}$ is that of the fitted GMM at the threshold which is set at the intersection of the two distributions. When $|\mu_1 - \mu_2|$ is large relative to $\sigma_1+\sigma_2$, $S_1$ is smaller, and the two peaks are more distant and separated. When $PD_{th}$ is small relative to $PD_{small\ peak}$, $S_2$ is small, and the two peaks are more intense. In these cases, it is easy to classify the signal points into two classes.

The algorithm for noise reduction and clustering is as follows. Note that GMM is used not only for the determination of the appropriate κ, but also for clustering the signal points.

1. Let K be the set of candidates of κ. Set lists $list_{s1}$ and $list_{s2}$ empty.
2. Repeat the following steps for all candidates of κ in K.
   a) Perform noise reduction for the signal using Kalman filter with parameter κ.
   b) Cluster the signal points into two levels using a GMM-based approach.
      i. Fit a two-component Gaussian Mixture Model (GMM) for the filtered signal. The component with the larger mean corresponds to the open state, and the other component corresponds to the closed state.
      ii. Estimate component-member posterior probabilities for signal points using the fitted GMM. These represent cluster membership scores. Each signal point is assigned to the component (the signal level) with the higher posterior probability.
   c) Compute the separations $S_1$ and $S_2$ for the fitted GMM. Append $S_1$ to $list_{S1}$, and $S_2$ to $list_{S2}$, respectively.
3. Let $t_1$ and $t_2$ be the values of κ corresponding to the minimum values in $list_{S1}$ and $list_{S2}$, respectively.
4. The appropriate κ is determined to be the smaller value of $t_1$ and $t_2$.

**The detrend algorithm of AI2**

Detrend refers to removing baseline drift. We use a Random Sample Consensus (RANSAC) algorithm (24) for the boundary points of the signal. The use of RANSAC makes the bias estimation more robust for outliers than the conventional least square method. Moreover, it is also robust for the signal that has two levels corresponding to open and closed states and transients. In this study, we targeted linear bias. The algorithm is as follows.

1. Repeat the following steps 100 times
   (a) Sample 0.05% of points from the signal points.
   (b) Compute the boundary points including the sampled points.
   (c) Estimate the line fitting the boundary points using RANSAC.
2. The slope of baseline drift is determined to be the line parameter that appears most frequently in 100 iterations.

**Generating datasets of pseudo-channel-currents**

We generated datasets that simulate ion channel currents with different SNRs as shown in Fig. 2. Time-series data consisting of closed (C) and open (O) states were first simulated using the QuB software (7). Channels are assumed to follow a two-state Markov process, and their transition between C and O states is governed by a continuoustime Markov process with transition rates $k_1$ and $k_2$ (*Eq.3*).

$$C \underset{k_2}{\overset{k_1}{\rightleftarrows}} O \text{ (Eq. 3)}$$

Four values (1, 10, 100, and 1000 $s^{-1}$) were examined for both $k_1$ and $k_2$, yielding 16 combinations of $k_1$ and $k_2$. For each combination, five time-series event data were simulated as the ground truth. Each data set contained $N= 5\times10^5$ points which correspond to a 20 s recording with a sampling frequency of 25 kHz.

Experimental or white Gaussian noise was then added to the ground truth. For the addition of experimental noise (Fig. 2 (b)), voltage sequences consisting of $V_{closed}$ (=0 V) and $V_{open}$, which corresponded to ground truth, were input to a 10 MΩ resister using the BATH terminal of the Patch-1U model cell (Molecular Devices) (10) from a function generator (multichannel systems STG4000s) controlled by the MC_stimulus II software. The resulting current signals were recorded for 20 s using an Axopatch 200B patch-clamp amplifier (Molecular Devices). The current signals were filtered at 2 kHz with an analogue low-pass Bessel filter, digitized at 25 kHz, and stored online using a data acquisition system (Digidata 1550B and pCLAMP 11). The SNR was defined as $(I_{open}-I_{closed})/I_\sigma$, where $I_{open}$ and $I_{closed}$ are the average currents recorded at $V_{open}$ and 0 V, respectively, and $I_\sigma$ is the standard deviation of the current at $V_{open}$. Computed channel current datasets were also generated by adding the white Gaussian noise to the ground truth (Fig. 2(a)). For each SNR, white Gaussian noise with the same $I_\sigma$ as the experiments were added to the C and O states using python 3.7. In total, we created 480 pseudochannel-current datasets, including those with experimentally recorded and computersimulated noise (Figs. S1-S6).

### Quantitative performance metrics

To quantify the performance of the method, two metrics were calculated: mean open time and error rate. Mean open time was obtained as an arithmetic mean of the observed dwell-time at the O state. The error rate was calculated by dividing $N_{fail}$ by $N_{all}$ where $N_{fail}$ was the number of data points in which the idealization failed to assign correctly and $N_{all}$ was all data points. In some analyses, we also counted the open event number per full trace (20 s). The accuracy (%) for the open event number and mean open time was defined as 100×(1-|I-R|/R), where I and R are the average values obtained from the idealization results and ground truth, respectively.

### Recording biological ion channel currents

Channel currents for gramicidin A (Sigma-Aldrich) and unidentified channels were recorded in bilayer lipid membranes (BLMs) that were formed across ~150-μm apertures in a 12.5-μm-thick Teflon film. The BLMs were prepared in respective recording solutions: 150 mM KCl and 10 mM HEPES (pH 7.4 with KOH) for gramicidin A; 120 mM KCl and 10 mM HEPES (pH 7.2 with KOH) for the unidentified channel. The lipid solutions used for BLM preparations were as follows: 5 mg/ml 1, 2dioleoyl-snglycero-3-phosphocholine (DOPC): cholesterol (Chol) = 4:1 (w/w) in chloroform/nhexane (1:1, v/v) for gramicidin A; 10 mg/mL asolection in n-hexane for the unidentified channel. For the incorporation of gramicidin A channels in BLMs, a methanolic solution of gramicidin A that had been diluted with its recording solution was added to both-side recording solutions (*cis* and *trans*) surrounding the BLMs. Proteoliposomes containing the unidentified channels were prepared in the following procedures. The open reading frame, which encodes a 699-aa polypeptide, including KAT1 with C-terminal 8-histidine tag (KAT1-HT), was amplified by PCR from the plasmid DNA, pDESTKAT1-HT2 (25),
with primers, 5′

GAGCTCAACAATGTCGATCTCTTGGACTC-3′ and 5′-

GGATCCTTAGTGATGGTGATGGTGATGGTGGTGCTCGAGTTCGGA-3′, and cloned into pTA2 (TArget Clone™; TOYOBO, Osaka, Japan), resulting in pTA2KAT1HT. A KAT1-HT cDNA fragment excised from the pTA2KAT1-HT with restriction enzymes, SacI and BamHI, was cloned into the SacI-BamHI sites of pEU3b (26). The resulting plasmids were used to generate mRNA *in vitro* with SP6 RNA polymerase, and CF protein synthesis was performed with the mRNAs and wheat germ extracts in the presence of asolectin-liposomes as previously described (27).

Transmembrane voltage was applied via two Ag/AgCl electrodes that were placed in the recording solutions. Applied potentials were defined with respect to the *trans* side, which was held at the ground. Transmembrane currents were recorded with an Axopatch 200B patch-clamp amplifier. The current signals were filtered with an analogue lowpass Bessel filter, digitized, and stored online using a data acquisition system (Molecular Devices Digidata 1440A using pClamp 10). The cut-off and sampling frequencies were 1 and 10 kHz, respectively.

### Analysis based on 50%-threshold-crossing and MDL

The 50%-threshold-crossing method was applied to the recorded currents as follows. The current datasets were first low pass filtered using an 8 poles Bessel filter. The cutoff frequencies were 500, 700, 1000, and 2000 Hz, when the larger value of $k_1$ and $k_2$ was 1, 10, 100, and 1000, respectively. Binomial Gaussian distribution fitting was performed for the filtered data. The threshold was determined as the midpoint of the two peaks of the distribution. Data points higher than the threshold were labeled as O, and a current level equal to or lower than the threshold was labeled C. MDL code is available on MATLAB (21). The parameter lmin, which is necessary for idealizing by MDL method, was set to 5.

## Results and Discussion

### Testing AI2 using pseudo-channel-current datasets

We first applied AI2 to examine pseudo-channel-current datasets that contained no baseline drift. The datasets were generated by adding white Gaussian or experimentally recorded noise currents to time-series data consisting of C and O states that were simulated using QuB (6, 7). The different levels of SNRs were examined: high (10.2), medium (5.14), and low (1.82). When $k_1$ and $k_2$ were 1, the datasets showed clear rectangular currents for both Gaussian or experimental noise (Figs. 3a and 3b).

The current exhibited infrequent transitions and long dwell-times for the O and C states. The AI2 was able to analyze the current data to yield idealization results that were perfectly matched with the ground truth. As summarized in Table 1, the error rate for assigning C or O states was less than 1 % for the current dataset with white Gaussian noise. Consequently, the open event number and mean open time were estimated with an accuracy in excess of 99.9%. Even for the dataset with experimental noise, the accuracy of the open event number and mean open time was higher than 94%.

When $k_1$ and $k_2$ were 100, the datasets showed frequent transitions between O and C states (Figs. 3c and 3d). There were more intermediate data points for $I_{open}$ and $I_{closed}$, which made the assignments of states more challenging. The idealization results obtained using AI2 were in good agreement with the ground truth for both types of noise. The error rate was still ~1 % for the datasets with white Gaussian noise, leading to the detection of open events with an accuracy of 98%. The mean open time estimated by AI2 was also highly accurate at 98%. In the datasets with the experimental noise, high-frequency noise was attenuated through a low pass filter on the amplifier and low frequency noise contributed more to current fluctuation (Fig. S7), resulting in the actual ion-channel recordings being essentially mimicked. Although the performance of the AI2 idealization was slightly lower compared with those for the datasets with white Gaussian noise, the idealization still recovered 94% of the open events with an error rate of less than 3%. The accuracy of the mean open time was as high as 94%.

**Estimating the appropriate κ of the Kalman filter**

AI2 is the algorithm that determines the appropriate κ during the iteration of Kalman filter noise reduction and GMM clustering based on the calculated parameters $S_1$ and $S_2$. We evaluated the validity of the determined κ in terms of mean open time, which is one of the most important parameters for describing the functions of ion channels. We calculated the mean open time from the idealized data, and the deviation from the theoretical value ($1/k_2$) (28) was quantified as α (= $k_2$×(mean open time)). When the mean open time obtained from the idealization results coincided with the theoretical value, α was 1. When the mean open time was estimated to be shorter or longer than $1/k_2$, α was lower or higher than 1, respectively.

Fig. 4 (a, b) shows the relationship between κ of the Kalman filter and distribution parameters ($S_1$ and $S_2$), and the calculated α (n=5). $S_1$ and $S_2$ are parameters that represent the degree of peak separation in the current distribution histogram along the horizontal axis (current) and the vertical axis (probability density), respectively. The search for the appropriate value for κ starts from $1.0\times10^{-1}$ to $1.0\times10^{-4}$. With a decrease in κ, $S_1$ gradually decreased, reaching a minimum value around κ of $10^{-2}$ and then increased. Similarly, $S_2$ also reached a minimum around a κ value of $10^{-2}$ and became unstable when κ was smaller than $10^{-3}$. Based on these results, AI2 determined the appropriate κ as $(9.65\pm0.68)\times10^{-3}$ (n=5, mean±standard deviation). Fig. 4 (c) shows an example of the idealization results for AI2 using the determined κ. The current histogram obtained from the filtered trace showed two clear peaks and AI2 provided appropriate binarization. The α of the idealized traces after the AI2 process was 1.03±0.04 (n=5). This result demonstrates that AI2 is capable of accurately estimating the mean open time through the automatic determination of the appropriate κ.

### Comprehensive evaluation of AI2 and comparison with 50%thresholdcrossing

To comprehensively evaluate the performance of AI2, α was obtained for datasets with white Gaussian or experimental noise at three SNR levels and visualized as heatmaps (Figs. 5-6). For comparison, the performance of 50%-threshold-crossing was also evaluated and plotted in the same manner. In the case of datasets with white

Gaussian noise, the idealization results obtained by both the AI2 and

50%thresholdcrossing agreed well with the ground truth for all datasets (Fig. 5). On the other hand, in the case of experimental noise (Fig. 6), AI2 showed a superior performance for 50%threshold-crossing. With 38 out of 48 combinations of transition rates and SNRs, the α value of AI2 was closer to 1 than that for 50%-threshold-crossing, indicating that AI2 performed better in 79% of the combinations. For example, when $k_1=1$, $k_2=10$, and SNR was low, the α of AI2 and 50%-threshold-crossing was 0.906±0.112 (n=5) and 0.216±0.028 (n=5), respectively. As shown in Fig. 7, AI2 detected 13 open events, exactly the same as that of ground truth, whereas

50%threshold-crossing assigned 74 open events. Due to the large number of false opens,

50%-threshold-crossing estimated a much shorter mean open time. The $S_1$ value (0.247±0.003) of AI2, which was calculated based on a current distribution histogram of the filtered traces, was smaller than that (0.367±0.002) for 50%-threshold-crossing ($p<0.00001$, n=5; two-tailed t-test).

The difference between the two peaks $|\mu_1-\mu_2|$, which correspond to the single-channel current, was estimated to be 3.43±0.01 pA for AI2 and 3.59±0.01 pA for 50%thresholdcrossing. There is little difference between these values because the data was acquired using equipment with an open circuit noise of 0.145pA rms (29). The much smaller peak variance ($\sigma_1$ and $\sigma_2$) of AI2 led to a smaller $S_1$ and better signal separation. Similarly, AI2 yielded a smaller $S_2$ value ($(2.97\pm0.74)\times10^{-4}$) than 50%thresholdcrossing ($0.0267\pm0.0017$) ($p<0.00001$, $n=5$; two-tailed t-test). In the case of 50%thresholdcrossing, the two peaks of the current distribution were not well separated, resulting in a heavy-tailed distribution. The threshold was set at the tail of the closed state, giving rise to an increase in false opens. In the case of AI2, the threshold was set at the intersection of the two probability density, leading to a smaller $S_2$ and better peak separation. These results demonstrate that AI2 is capable of automatically adjusting noise reduction to achieve adequate peak separation without user input, leading to better idealization performance than the conventional 50%-threshold-crossing.

**Verification that AI2 removes baseline drifts and application to actual biological channel currents**

In actual electrophysical recordings, ion channel currents often include a baseline drift. Detrend, i.e., removing baseline drift, is another important issue in the analysis of channel currents. We examined the detrend function of AI2 using experimental datasets with a linear baseline drift. Fig. 8 illustrates current traces before and after removing baseline drift and examples of idealization results in blue and green areas. In the original trace (Fig. 8 (a)), the current level increased linearly by 6.90 pA in 20 seconds. AI2 estimated that a baseline drift of a 6.90±0.41 pA increment in 20 seconds (n=5) was superimposed on the current signal. After detrend was completed, the current traces showed no baseline drift, and the $\alpha$ obtained after idealization was 1.08±0.02 (n=5), representing an accurate estimation of the mean open time. This shows that the detrend function of AI2 efficiently removes drift automatically.

Finally, we evaluated the applicability of AI2 for actual biological ion channel currents using three different ion channels: gramicidin A, the human cardiac sodium ($Na_v1.5$) channel, and unidentified channels. These ion channel currents were idealized using AI2 and compared with 50%-threshold-crossing and MDL idealization. As shown in Fig. 9, binary idealization results were obtained with AI2 and 50%-threshold-crossing. Although the single channel current of the gramicidin A channel was small (~1 pA), AI2 successfully distinguished open and closed states without being affected by the low SNR (Fig. 9 (a)). In the case of $Na_v1.5$ currents, the channel activities were evoked by using a voltage protocol with a short prepulse to –200 mV, followed by a step to +200 mV (Fig.

9 (b)). Application of the voltage step induced a transient current, and consequently, the $Na_V$ channel currents were superimposed on a baseline drift. AI2 was more robust for baseline drift than 50%-threshold-crossing and successfully detected the opening and closing events irrespective of the baseline drift, which, in the past, has been an obstacle to ion channel analysis. These results demonstrate that AI2 can automatically perform idealization, even when current data included unstable baselines. We also applied AI2 to ion channels whose function has not yet been identified. As shown in Fig. 9 (c), AI2 could idealize the currents of unidentified channels without *a priori* knowledge of the gating model. This demonstrates that AI2 is highly robust for processing actual electrophysiological data that contains various noise, baseline drifts, and channel kinetics.

## Conclusions

We developed the AI2 algorithm that allows the automatic and model-free idealization of ion channel currents containing complex noise and baseline drift which are often observed in actual electrophysiological measurements. AI2 can simultaneously perform noise reduction and clustering in the iteration of Kalman filtering and GMM fitting. By using pseudo-ion-channel-current datasets, it was demonstrated that AI2 can be used to determine the most appropriate κ of a Kalman filter and the idealization results were in good agreement with theoretical values. Although both AI2 and 50%thresholdcrossing are based on a combination of filtering and thresholding, AI2 enables the automatic determination of the filter constant that yields the best separation between O and C states without the need for the labor-intensive processes that are involved in determining the optimal filter setting. We also applied AI2 to three different electrophysiological currents and confirmed that AI2 enabled unsupervised idealization despite the complex noise, baseline drifts, and unidentified gating kinetics of the target channel. AI2 is highly robust and is applicable for use in assessing actual electrophysiological data. An algorithm for removing linear baseline drift in the detrend processing of AI2 was included as the first demonstration. In actual electrophysiological recordings, however, channel currents are often superimposed on more complex baseline drift. Further investigations will be needed for the estimation and elimination of higherorder baseline drifts. Although the present method is proposed for the analysis of singlechannel currents, it will be applicable to other types of time-series binary data, including those obtained with single-molecule fluorescent imaging.

## Author Contributions

M.S. conceived the study, designed and performed the experiments, analyzed the data, and wrote the manuscript. M.H. designed the research, contributed analytic tools, and wrote the article. M.K. performed the experiments. K.S. and Y.T. purified the samples. H.Y. designed the experiments. A.H.-I. designed the research and wrote the article. All authors discussed the results and commented on the manuscript.

## Declaration of Interests

The authors declare no competing interests.

## Acknowledgments

This work was supported by Grants-in-Aid for the scientific research from the Japan Society for the Promotion of Science (JSPS), Japan (19H00846, 20K20550, 21H05164, and 21K14505). M.S. is supported by Research Fellowships for Young Scientists from JSPS (22J13311) and the WISE program for Artificial Intelligence and Electronics at Tohoku University. Some of the equipment used in this research was manufactured by Kento Abe, a technical staff member in the machine shop division of Fundamental Technology Center, Research Institute of Electrical Communication, Tohoku University.

Table 1. Comparison of AI2 idealization results with ground truth

| $k_1=k_2=1$ (1/$k_2$=1.000 s) | | | |
|---|---|---|---|
| | | Reference value* | AI2 |
| White Gaussian noise Fig. 3(a) | open event number (per 1 × $10^5$ points) | 9.20 ± 2.71 | 9.20 ± 2.71 |
| | mean open time (s) | 0.954 ± 0.188 | 0.954 ± 0.187 |
| | error rate (%) | 0 | 0.118 ± 0.022 |
| Experimental noise Fig. 3(b) | open event number (per 1 × $10^5$ points) | 9.20 ± 2.71 | 9.00 ± 2.61 |
| | mean open time (s) | 0.954 ± 0.188 | 0.902 ± 0.219 |
| | error rate (%) | 0 | 0.262 ± 0.082 |
| $k_1=k_2=100$ (1/$k_2$=10 ms) | | | |
| | | Reference value* | AI2 |
| White Gaussian noise Fig. 3(c) | open event number (per 1 × $10^5$ points) | 978 ± 16 | 959 ±17 |
| | mean open time (ms) | 10.2 ± 0.1 | 10.4 ± 0.17 |
| | error rate (%) | 0 | 1.19 ±0.02 |
| Experimental noise Fig. 3(d) | open event number (per 1 × $10^5$ points) | 978 ± 16 | 921 ± 17 |
| | mean open time (ms) | 10.2 ± 0.1 | 10.8 ± 0.2 |
| | error rate (%) | 0 | 2.73 ± 0.26 |

*Reference value was calculated based on the ground truth. Values are presented as mean±standard deviation.

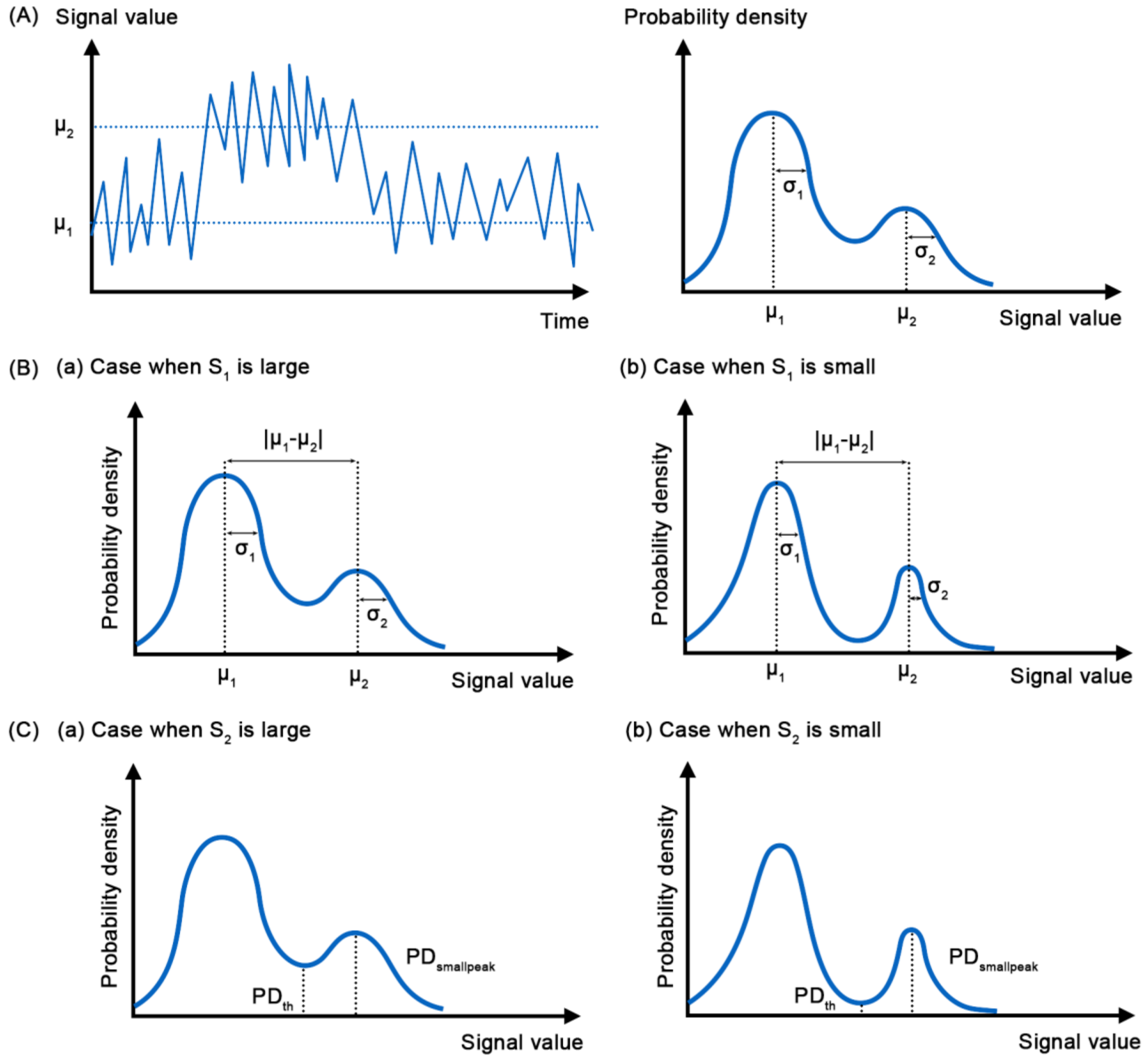

Figure 1. Principle of AI2 for the automatic determination of the degree of noise reduction. (A) Schematic of raw current signals and probability density generated from a histogram of the signals. (B) The relation between $S_1$ and the distribution of probability density. (C) The relation between $S_2$ and the distribution of probability density.

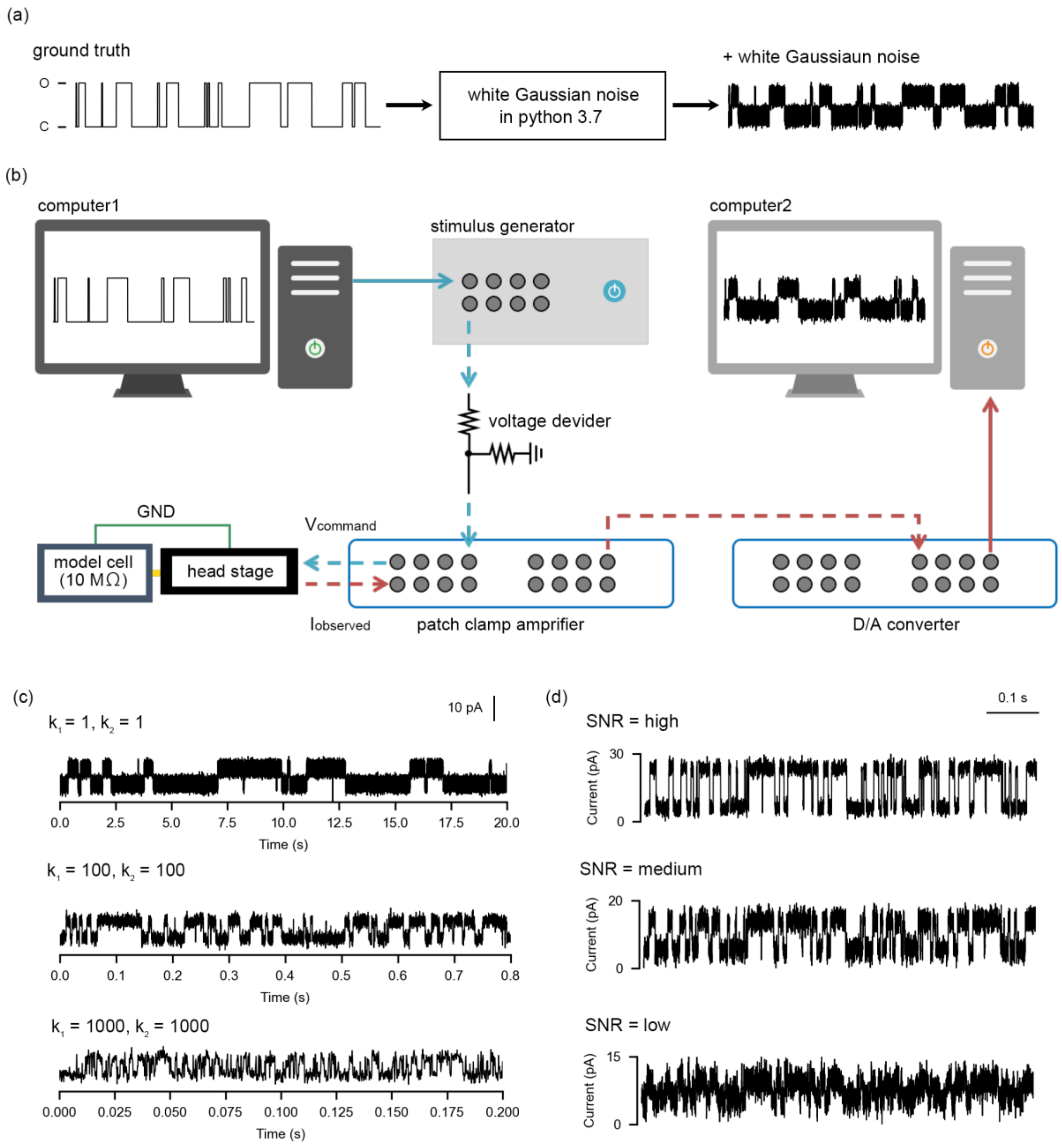

Figure 2. Workflow diagram for generating datasets that mimic ion channel currents. (a) Generation of datasets with white Gaussian noise. (b) Generation of experimental datasets that include electrophysiological recording noise. (c) Examples of experimental datasets with different transition rates $k_1$ and $k_2$ at SNR of 5.14 (medium). Top: $k_1=1$ and $k_2=1$, middle: $k_1=100$ and $k_2=100$, and bottom: $k_1=1000$ and $k_2=1000$. (d) Examples of experimental datasets with different SNRs at $k_1=k_2=100$. Top: high SNR (10.2), middle: medium SNR (5.14), and bottom: low SNR (1.82).

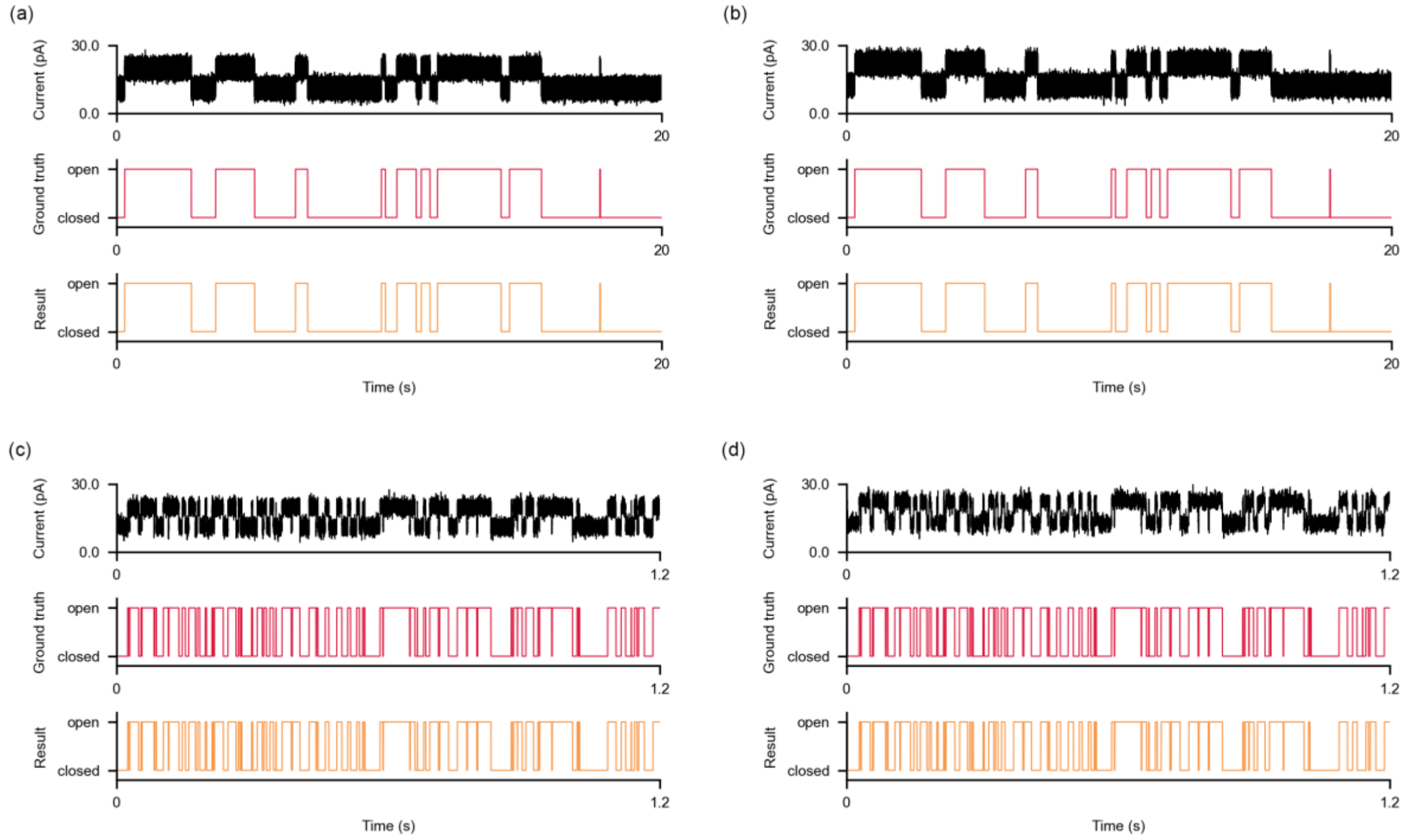

Figure 3. Examples of AI2 idealization results at an SNR of 5.14 (medium). The black lines show pseudo-ion-channel-currents (top), the red lines show ground truth (middle), and the orange lines show AI2 idealization results without a detrend function (bottom). (a, c) white Gaussian noise, (b, d) experimental noise. (a, b) $k_1$= $k_2$=1, (c, d) $k_1$= $k_2$=100.

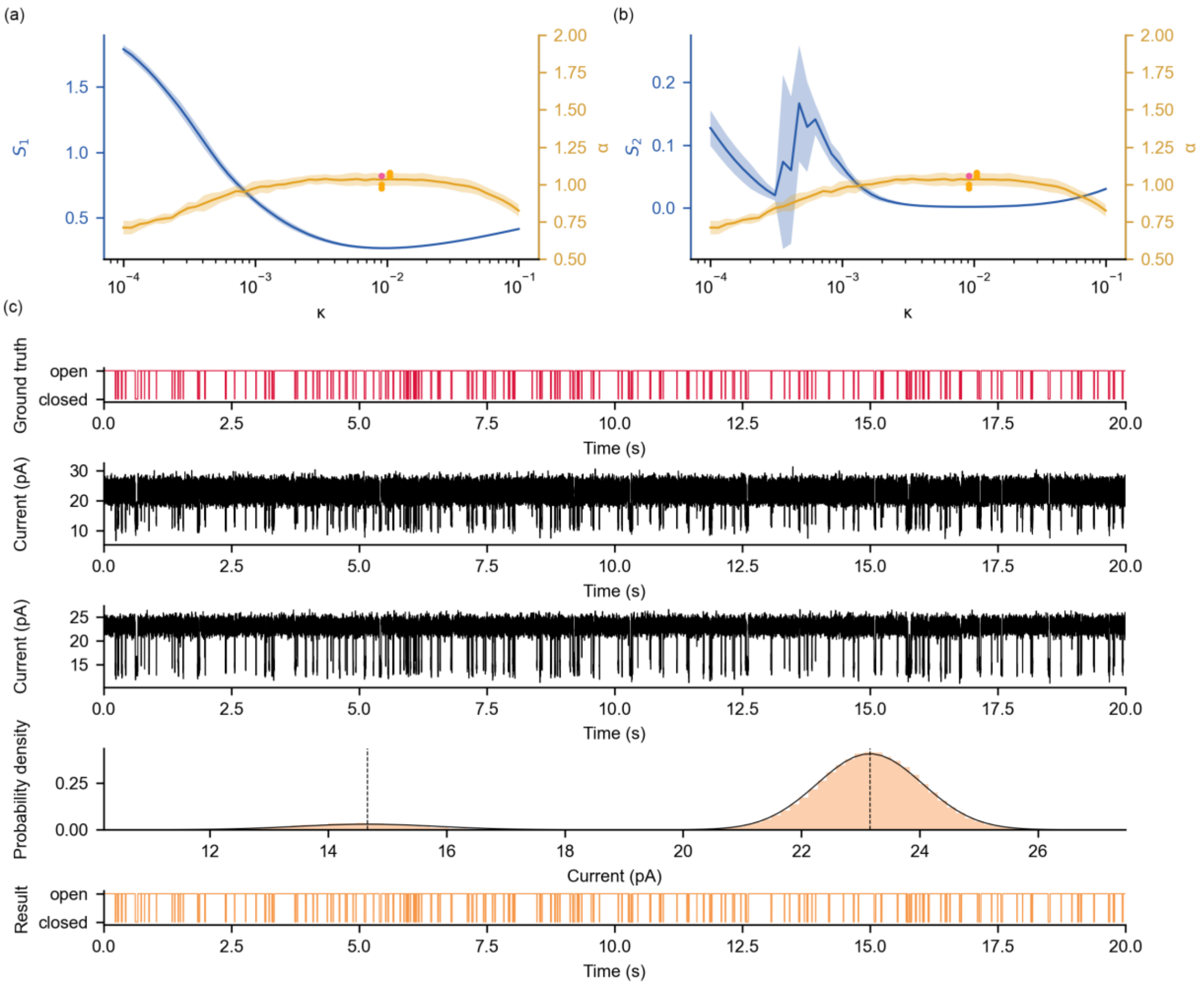

Figure 4. An example of the automatic determination of the Kalman filter parameter κ and idealization results by AI2 without a detrend function ($k_1$ =100, $k_2$ =10, medium SNR, and experimental noise). (a) The relationship between κ and $S_1$ (blue line), and κ and α (orange line). (b) The relationship between κ and $S_2$ (blue line), and κ and α (orange line). (a, b) The circles indicate the determined κ and the resulting α. Lines and shadows indicate means and standard deviations of 5 data, respectively. (c) Process flow of AI2 idealization for data marked with red circles in (a, b). From top to bottom: ground truth, trace of pseudo-ion-channel currents, trace after passing through the Kalman filter, histogram of the filtered currents, and idealization results.

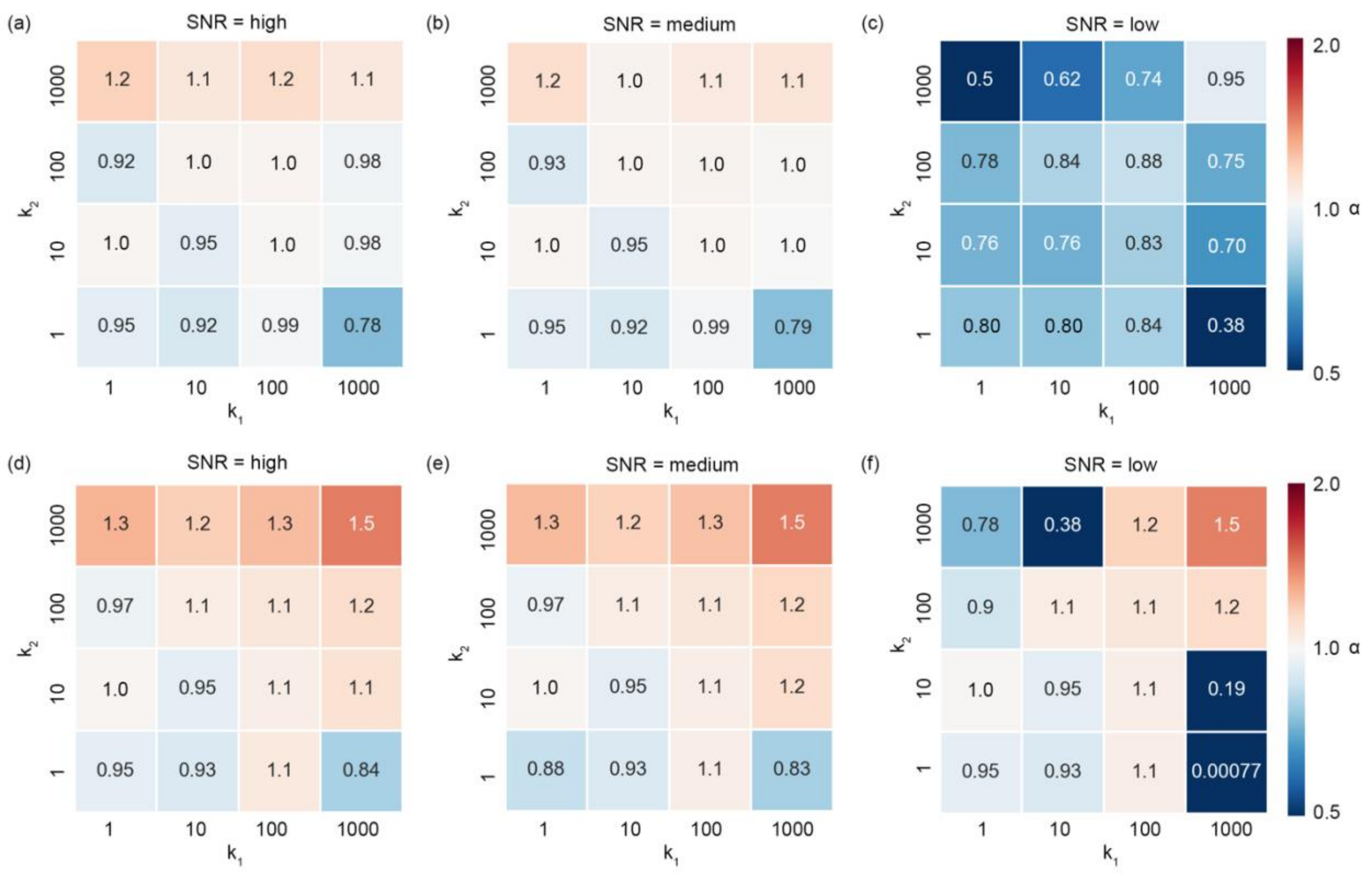

Figure 5. Comprehensive comparison of α between (a-c) AI2 (detrend function off) and (d-f) 50%-threshold-crossing for datasets with white Gaussian noise. The α ($k_2$×mean open time) is illustrated as heatmaps. Numerical values represent the mean of the results (n=5). (a, d) high SNR, (b, e) medium SNR, and (c, f) low SNR.

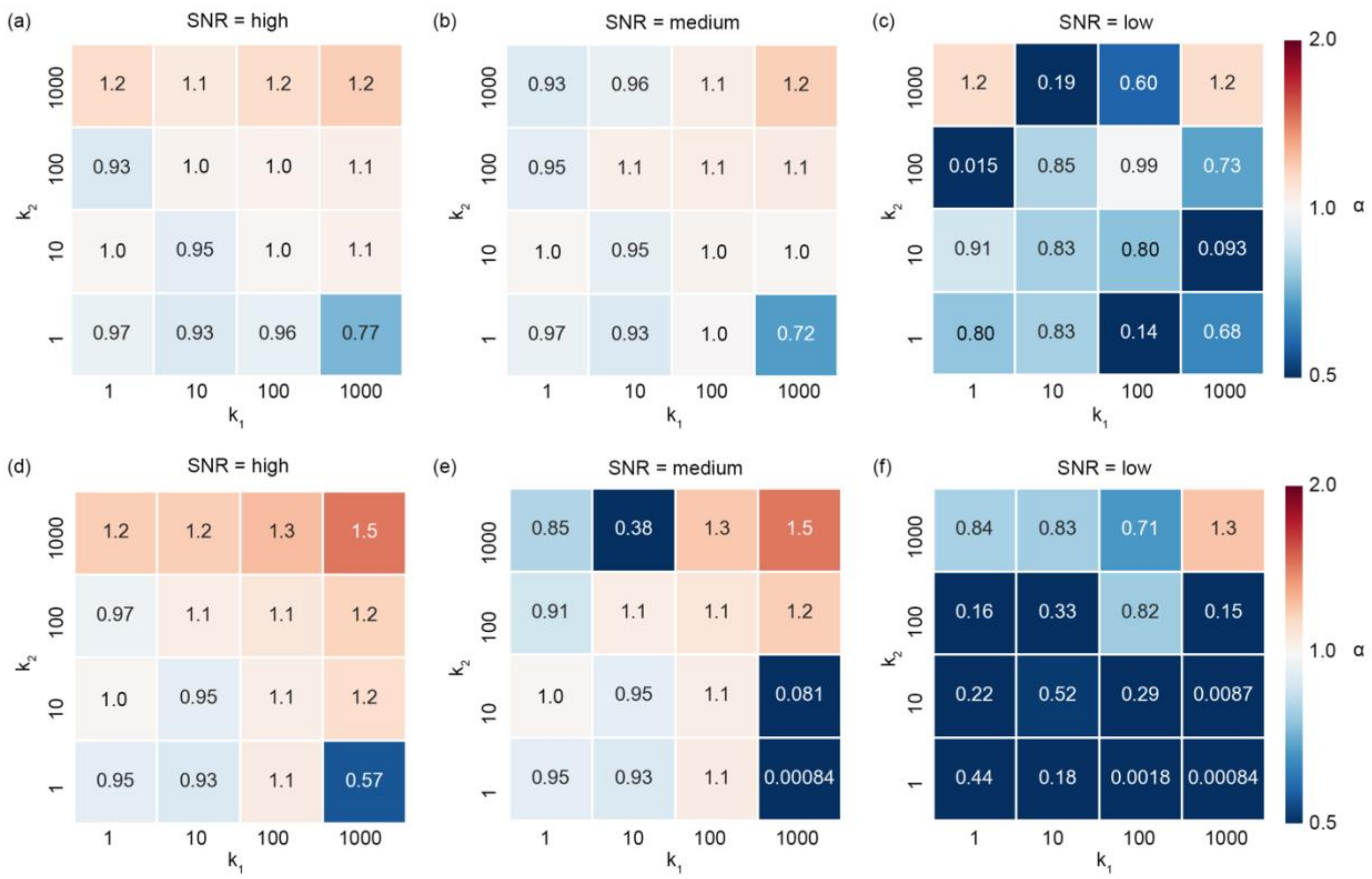

Figure 6. Comprehensive comparison of α between (a-c) AI2 (detrend function off) and (d-f) 50%-threshold-crossing for experimental datasets. The α is illustrated as heatmaps. Numerical values represent the mean of the results (n=5). (a, d) high SNR, (b, e) medium SNR, and (c, f) low SNR.

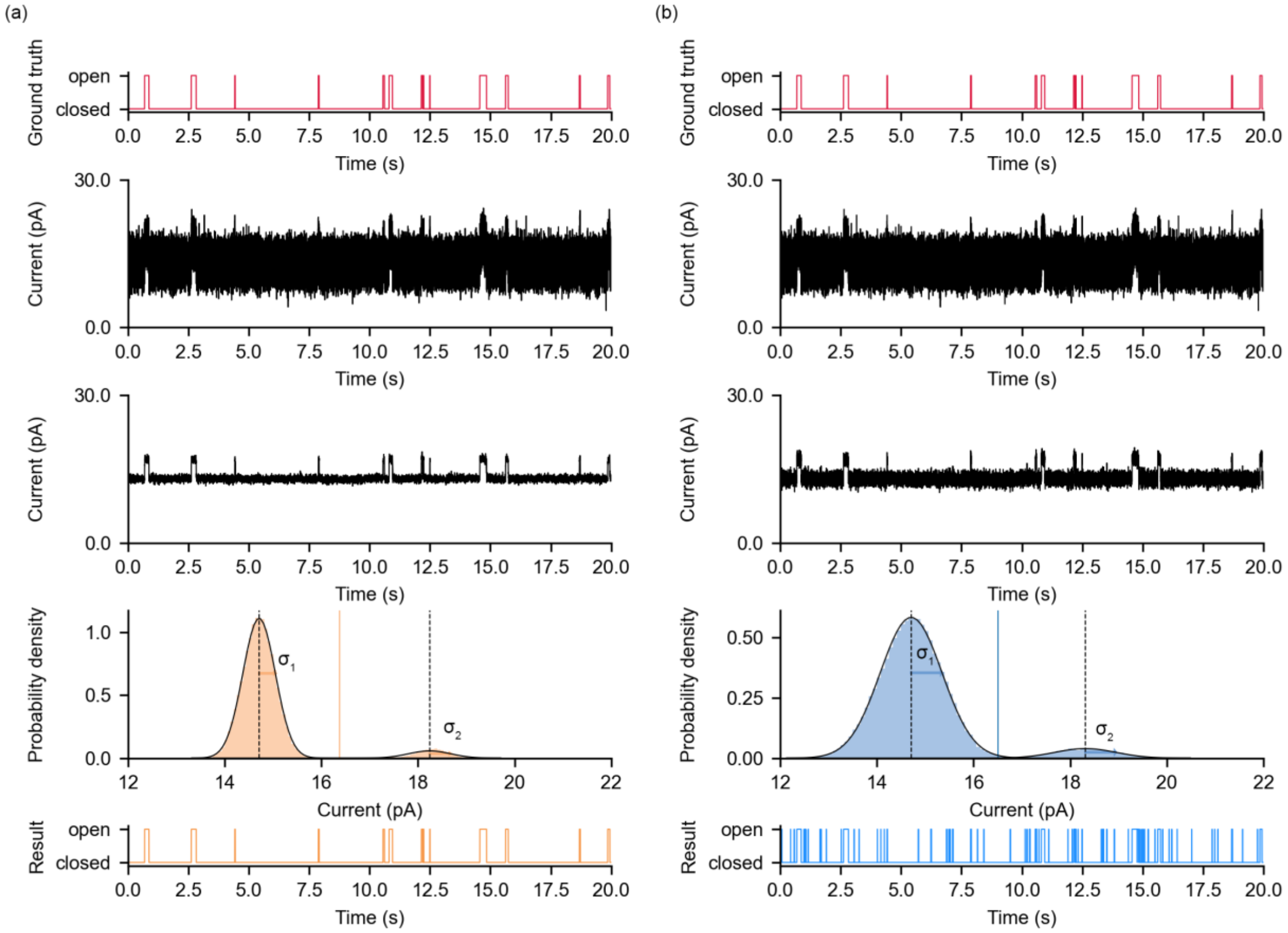

Figure 7. Process flow of (a) AI2 idealization without a detrend function and (b) 50%threshold-crossing ($k_1$= 1, $k_2$ = 10, low SNR, and experimental noise). From top to bottom: ground truth, trace of pseudo-ion-channel currents, trace after passing through the Kalman filter, histogram of the filtered currents, and idealization results.

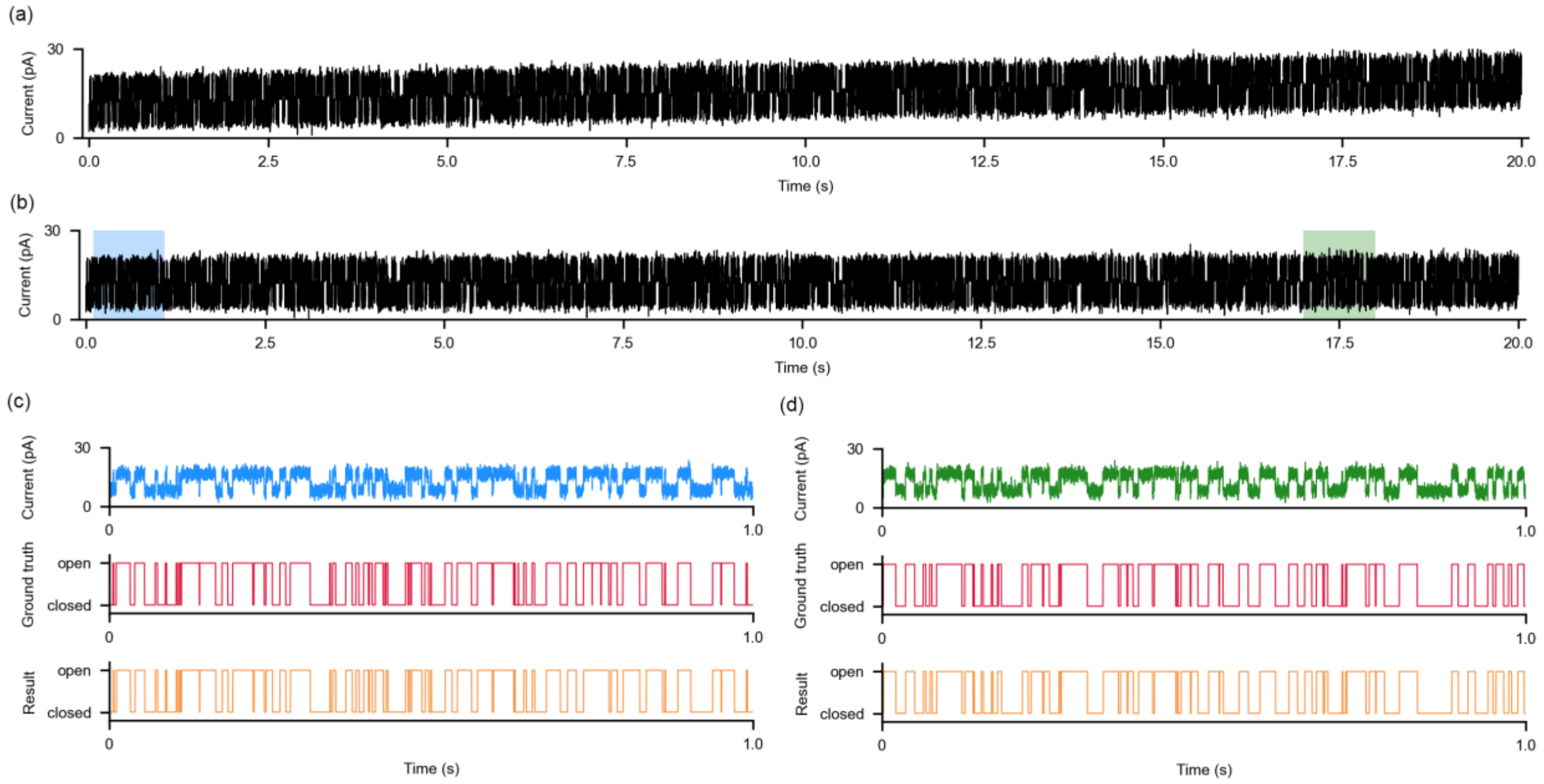

Figure 8. Idealization results of AI2 with detrend function for experimental datasets containing a baseline drift ($k_1$= $k_2$=100 medium SNR, and experimental noise). (a) Raw currents. (b) Current trace after detrend. (c, d) (top) Zoomed view of current traces highlighted in (c) blue and (d) green regions in (b), (middle) ground truth, (bottom) idealization results.

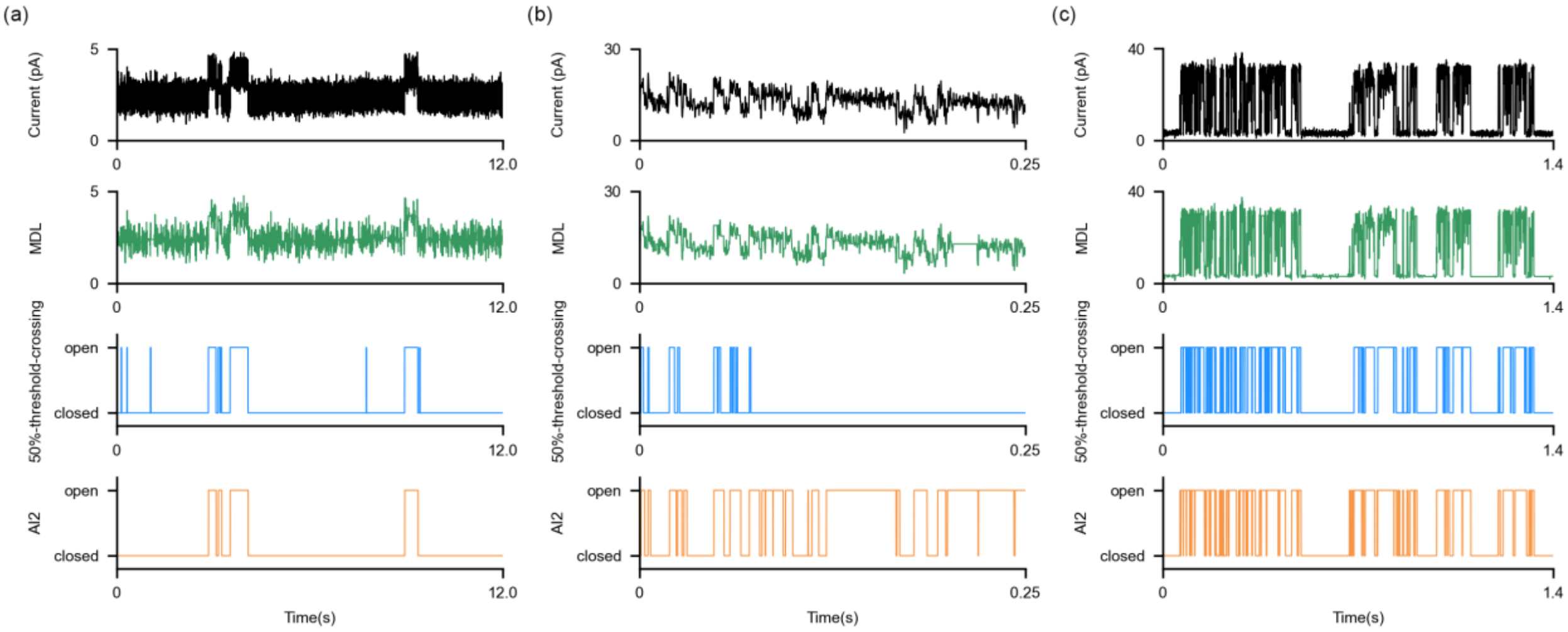

Figure 9. Comparison of idealization results for three kinds of ion channel currents. The black line represents channel currents. Green, blue, and orange lines represent the idealization results of MDL, 50%-threshold-crossing, and AI2 with a detrend function, respectively. (a) Gramicidin A channel currents recorded at 100 mV. (b) $Na_V$1.5 current data from ref. 30 which was recorded at 200 mV after a 10 ms prepulse of -200 mV. (c) Unidentified channel currents recorded at 100 mV. The cut-off frequency of the 50%threshold-crossing low pass filter was set at (a) 700 and (b, c) 1000 Hz.

# Supporting Material

# Model-Free Idealization: Adaptive Integrated Approach for Idealization of Ion Channel Currents (AI2)

*Madoka Sato[1], Masanori Hariyama[2*], Komiya Maki[3], Kae Suzuki[4,5], Yuzuru Tozawa[4], Hideaki Yamamoto[3], Ayumi Hirano-Iwata[1,3,6*]*

*[1]Graduate School of Biomedical Engineering, 2-1-1 Katahira, Aoba-ku, Sendai-shi, Miyagi 980-8577, Japan*

*[2]Graduate School of Information Sciences, Tohoku University, 6-3-09, Aoba, Aramaki, Aoba, Sendai, 980-8579, Japan*

*[3]Laboratory for Nanoelectronics and Spintronics, Research Institute of Electrical Communication, Tohoku University, 2-1-1 Katahira, Aoba-ku, Sendai-shi, Miyagi 980-8577, Japan*

*[4]Graduate School of Science and Engineering, Saitama University, 255 Shimo-Okubo, Sakura-ku, Saitama-shi, Saitama 338-8570, Japan*

*[5]Epsilon Molecular Engineering, Inc., Rm208, Research bldg. of Open Innovation Center in Saitama university, 255 Shimo-okubo, Sakura-ku, Saitama city, Saitama 338-8570 Japan*

*[6]Advanced Institute for Materials Research (WPI-AIMR), Tohoku University, 2-1-1 Katahira, Aoba-ku, Sendai-shi, Miyagi 980-8577, Japan*

**Correspondence: masanori.hariyama.c3@tohoku.ac.jp, ayumi.hirano.a5@tohoku.ac.jp*

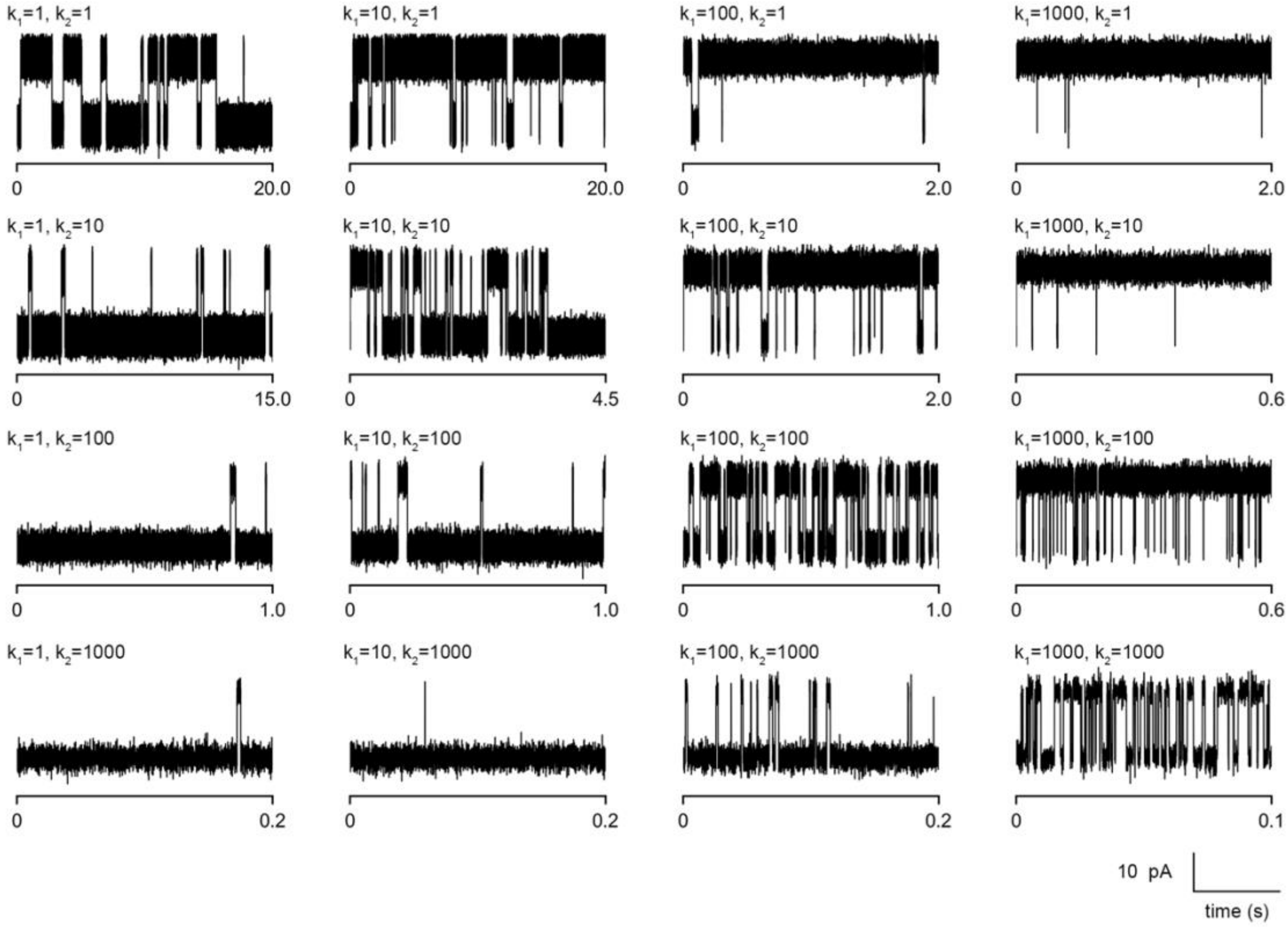

Figure S1. Examples of computational pseudo-channel-currents with high SNR.

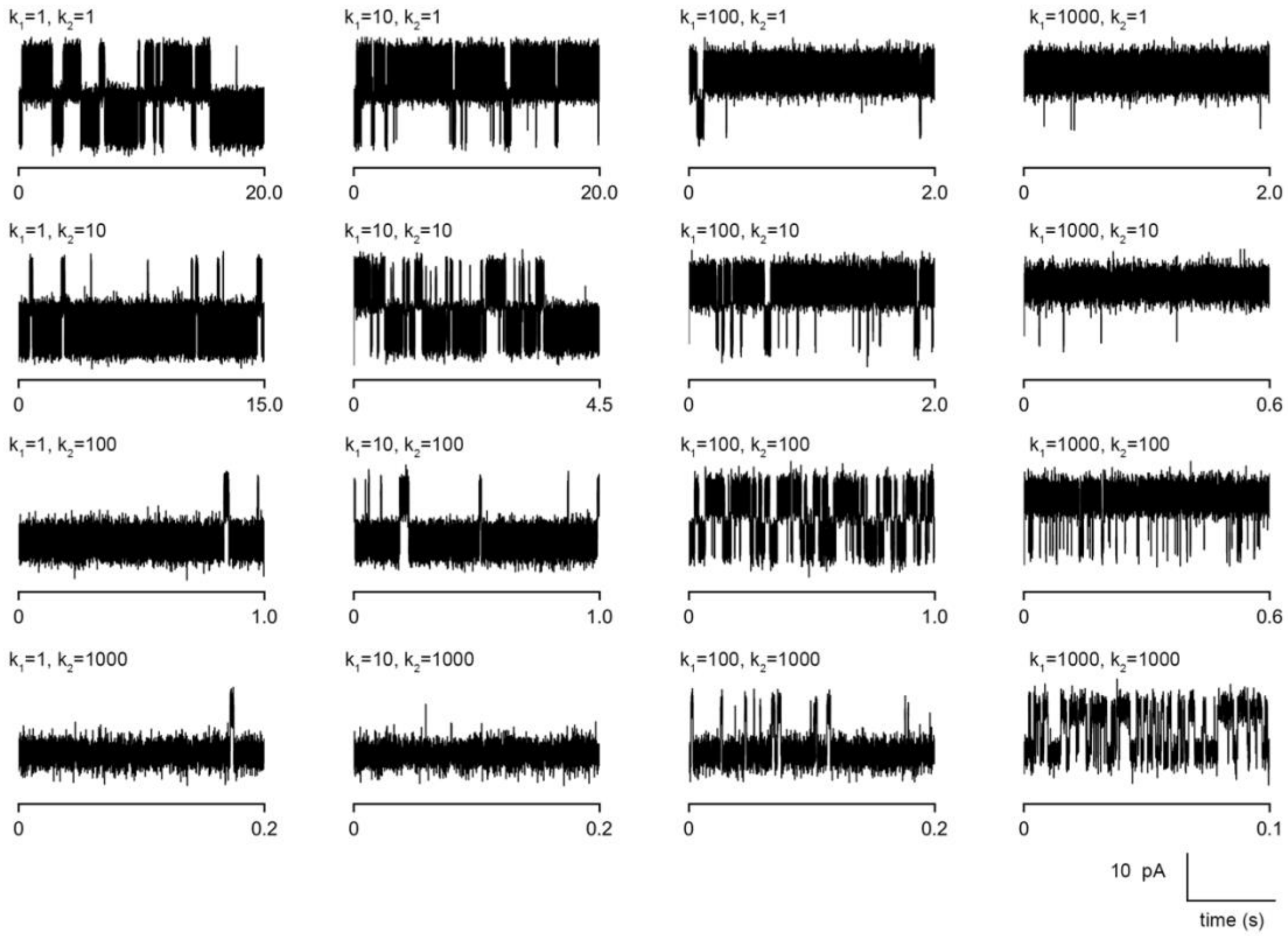

Figure S2. Examples of computational pseudo-channel-currents with medium SNR.

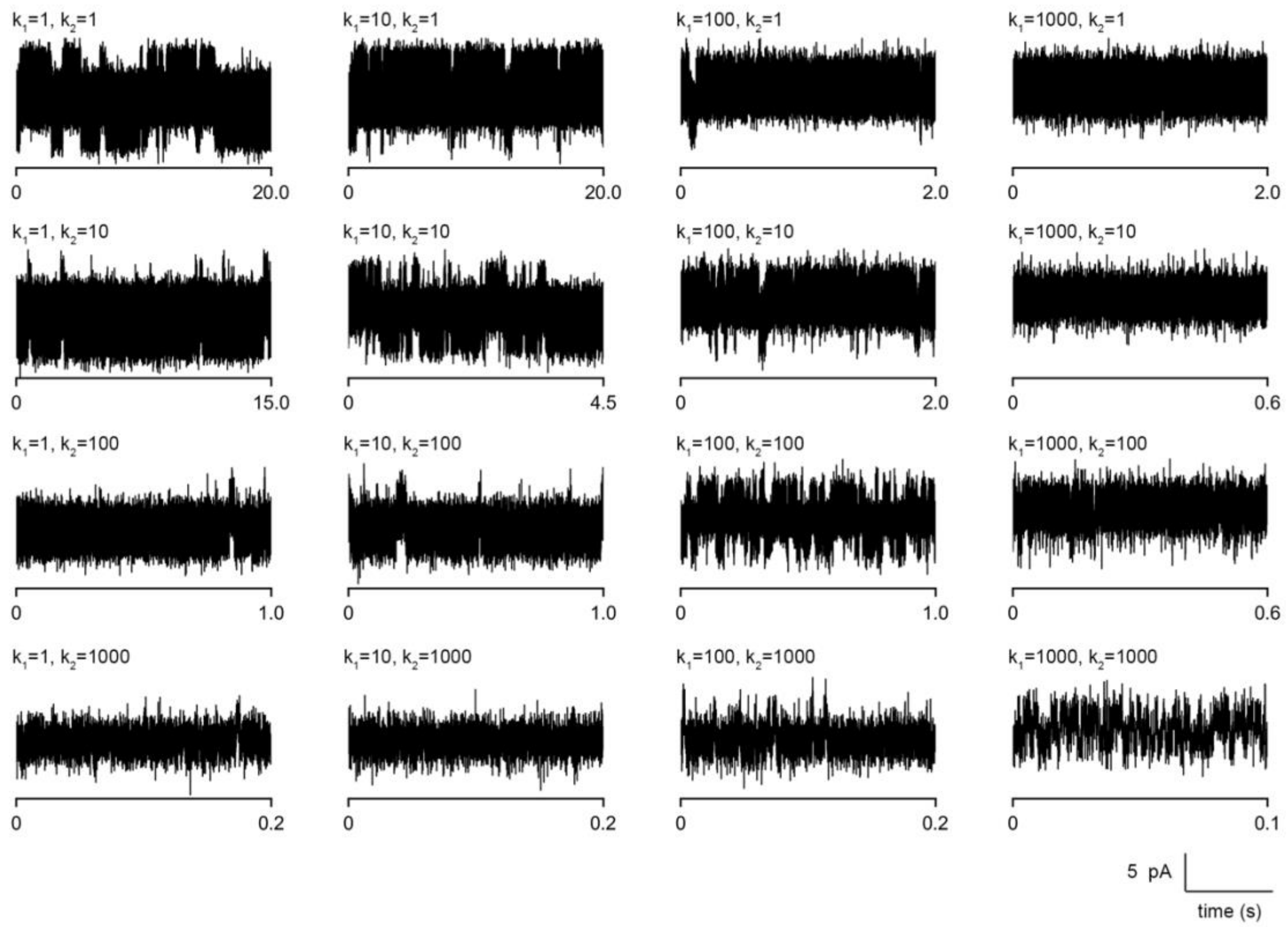

Figure S3. Examples of computational pseudo-channel-currents with low SNR.

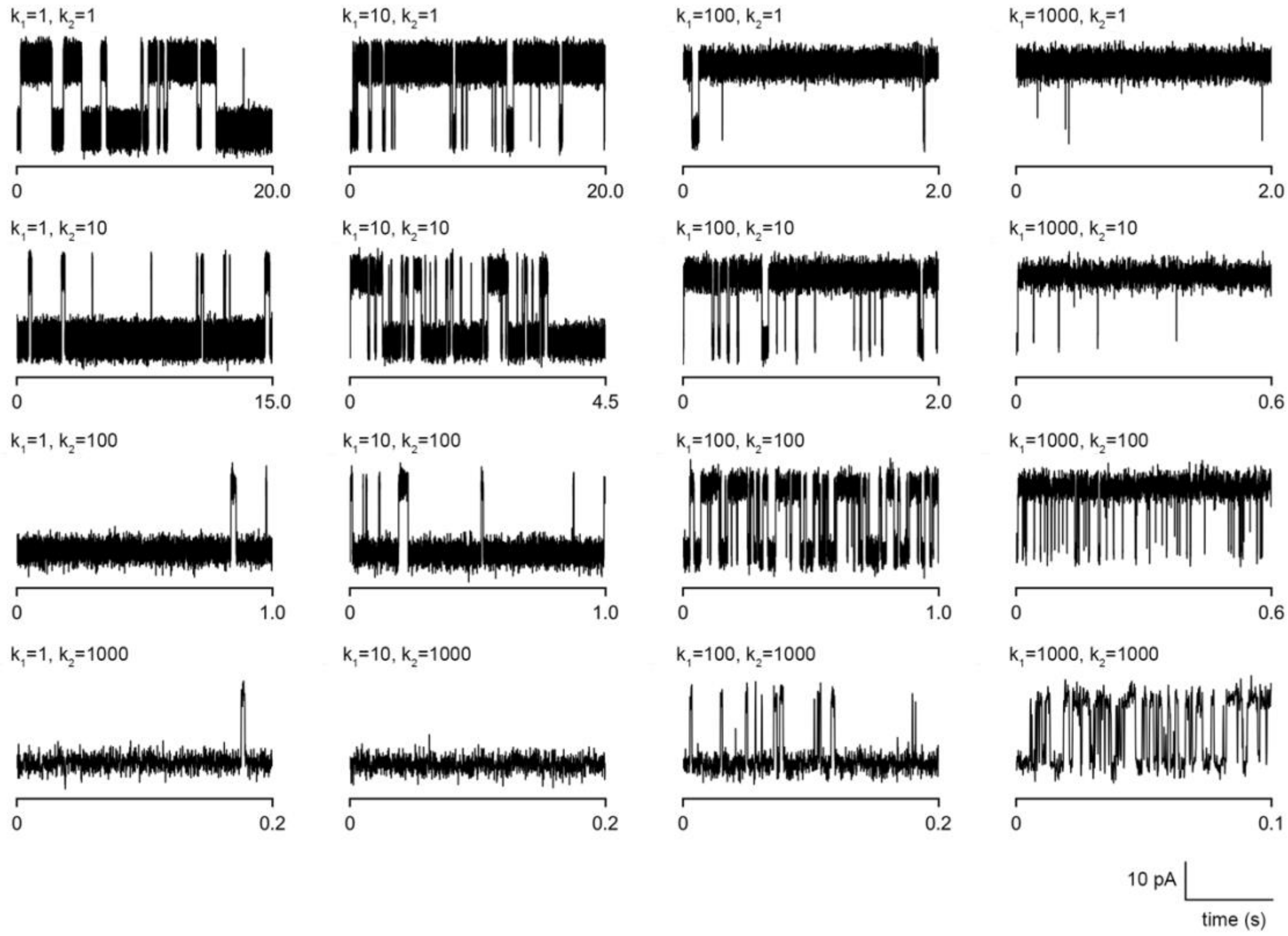

Figure S4. Examples of experimental pseudo-channel-currents with high SNR.

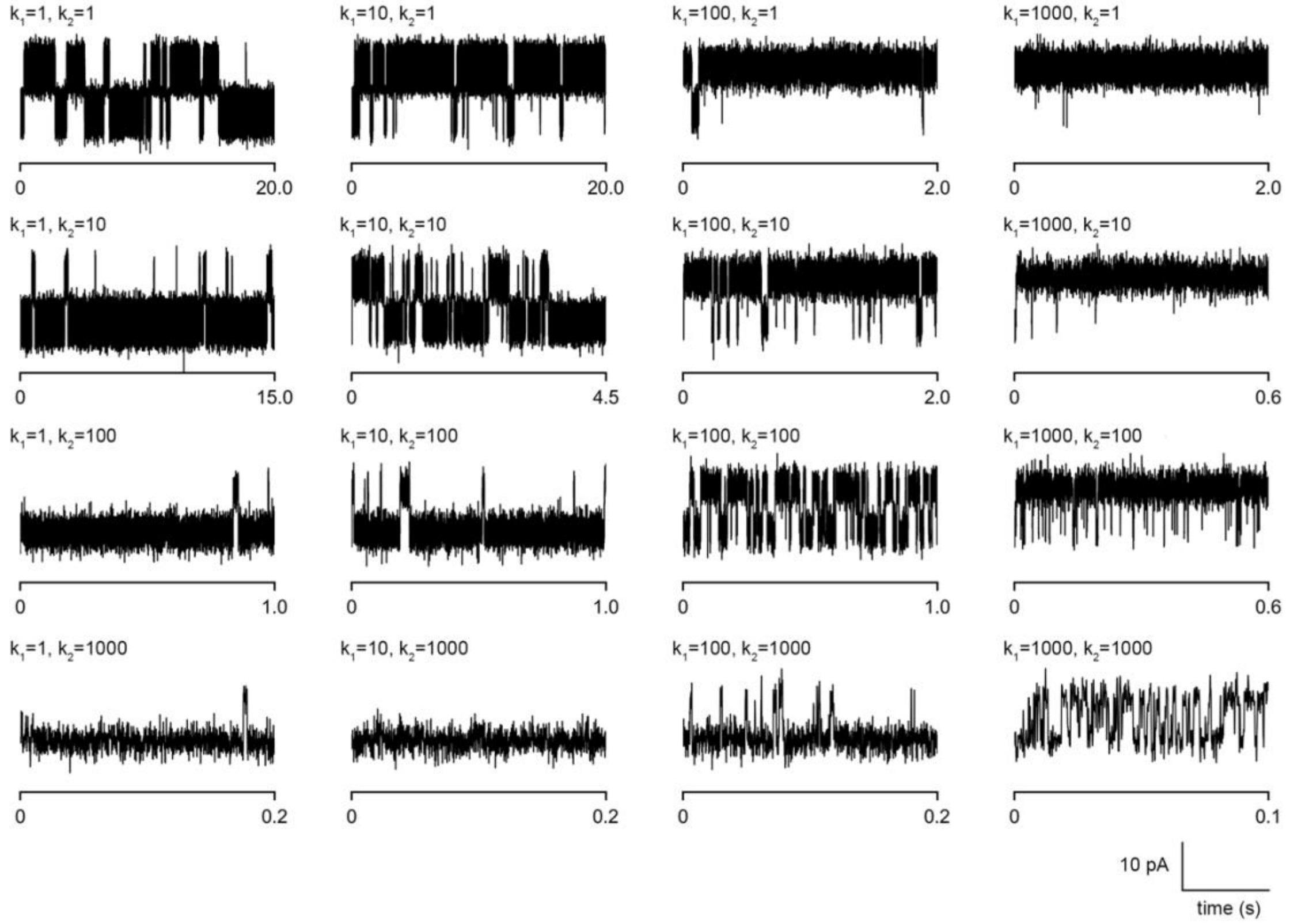

Figure S5. Examples of experimental pseudo-channel-currents with medium SNR.

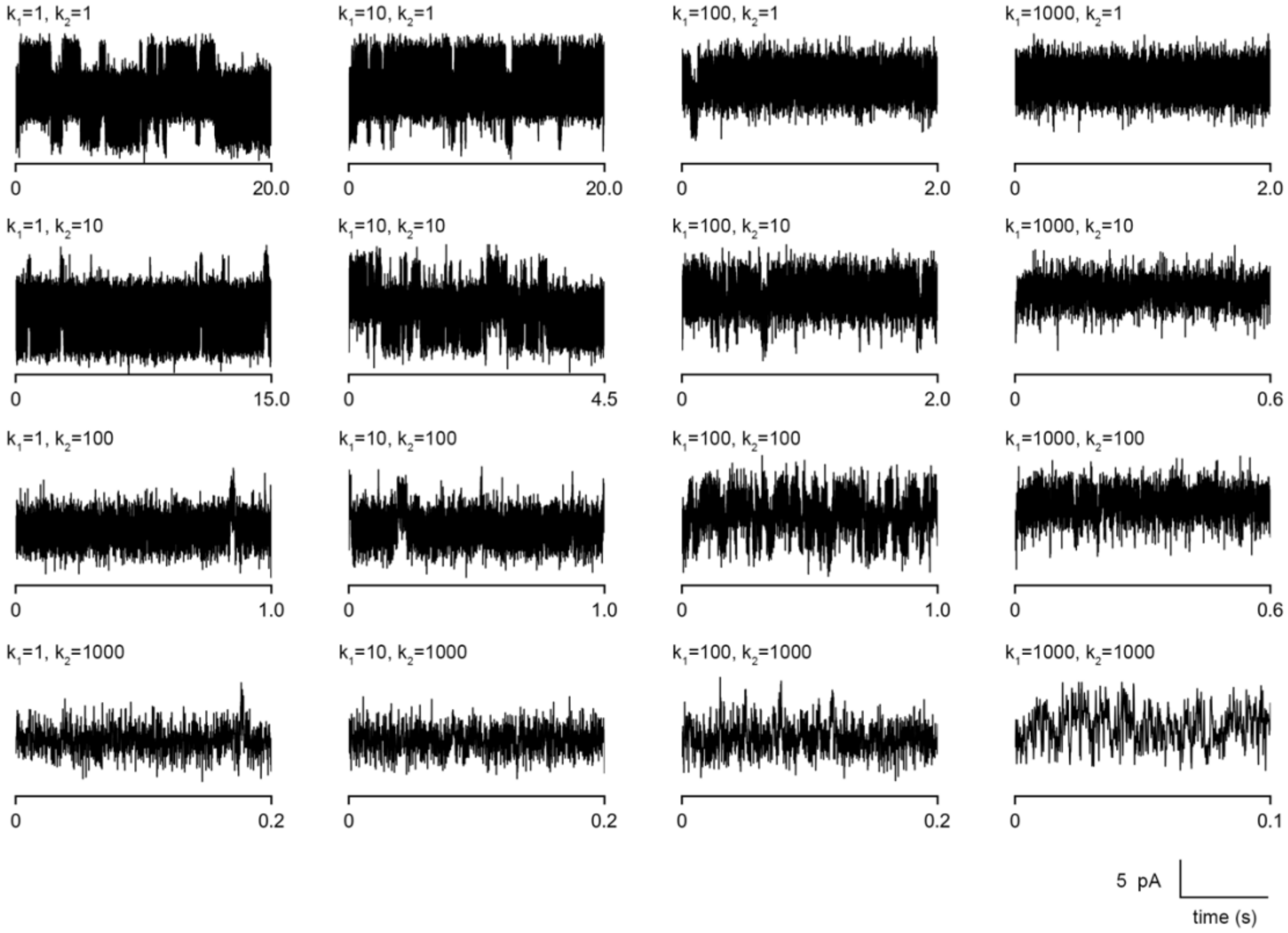

Figure S6. Examples of experimental pseudo-channel-currents with low SNR.

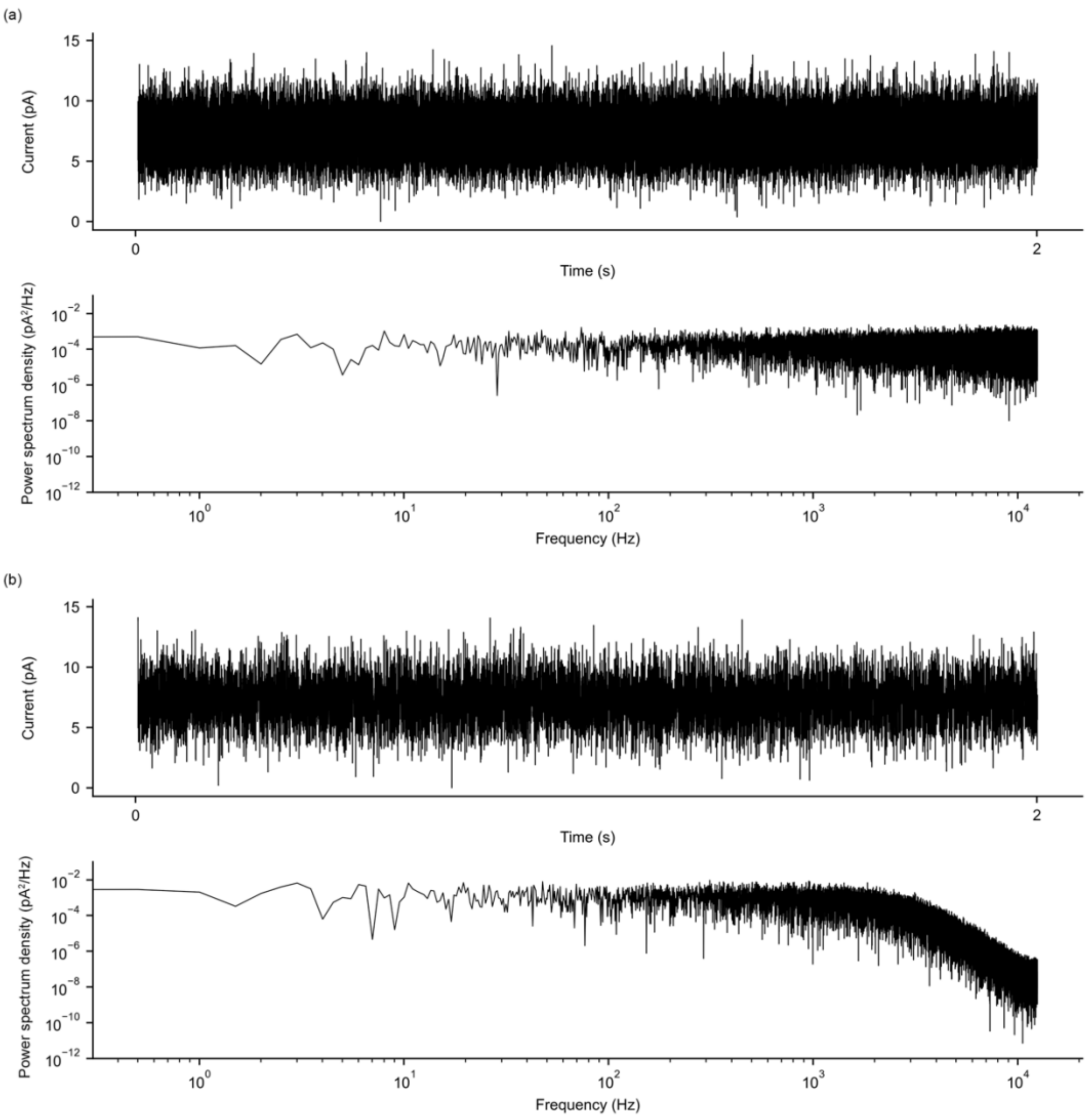

Figure S7. Current waveforms and their power spectrum density of the datasets with (a) white Gaussian noise and (b) experimental noise.